\documentclass[12pt]{article}

\usepackage{graphicx}
\usepackage{epsf}
\usepackage{amsmath, amssymb}

\begin{document}

\begin{titlepage}
\begin{center}

\hfill UT-10-21\\
\hfill TU-877\\

\vspace{1.5cm}
{\Large\bf Squark Mass Measurement in the Long-lived Stau Scenario at the LHC}

\vspace{1cm}
{\large Takumi Ito}

\vspace{1cm}

{\em
Department of Physics, University of Tokyo,
Tokyo 113-0033, JAPAN}

\vskip 0.07in

{\em
Department of Physics, Tohoku University,
Sendai 980-8578, JAPAN}

\vspace{1cm}
\abstract{
The long-lived stau scenario is an interesting possibility at the LHC.
We study squark mass measurements in this scenario; in particular, we show that left- and right-handed squark masses are both measurable.
In SUSY events, multiple jets are expected, which become a source of combinatorial backgrounds.
In order to reduce such backgrounds a hemisphere analysis is applied, and we discuss mass measurements of squarks in decay modes of $\tilde{q}_{L}$ into $q \tilde{W}^{0}$ or $q^{\prime} \tilde{W}^{\pm}$ and also $\tilde{q}_{R}$ into $q \tilde{B}$.
%%(PLAIN TEXT VERSION)
%\\
%The long-lived stau scenario is an interesting possibility at the LHC.
%We study squark mass measurements in this scenario; in particular, we show that left- and right-handed squark masses are both measurable.
%In SUSY events, multiple jets are expected, which become a source of combinatorial backgrounds.
%In order to reduce such backgrounds a hemisphere analysis is applied, and we discuss mass measurements of squarks in decay modes of left-handed squark into quark + (neutral/charged) Wino and also right-handed squark into quark + Bino.
}

\end{center}
\end{titlepage}

\renewcommand{\thepage}{\arabic{page}}
\setcounter{page}{1}
\renewcommand{\thefootnote}{\#\arabic{footnote}}
\setcounter{footnote}{0}

The CERN LHC experiment is now operating and seeking for the physics beyond the standard model.
The promising candidate of the new physics is the low-energy supersymmetry (SUSY),
and many studies are devoted for the discovery or the verification of the minimal supersymmetric standard model (MSSM) at the LHC.
However, such studies should strongly depend on the details of the MSSM.
For example, if some charged superparticle has a long lifetime enough to escape from the detector, the event topology of SUSY signals becomes quite different from the one in the conventional studies assuming the existence of stable invisible particle.

We pay particular attention to the case with the long-lived stau.
The lightest superparticle (LSP) is stable under the $R$-parity conservation and thus important for phenomenology.
For collider study, however, it is more important what the LSP in the MSSM sector (MSSM-LSP) is.
This is because the MSSM-LSP would not decay into the LSP inside the detector if the lifetime of the MSSM-LSP is long enough.
In a variety of the SUSY breaking models the lightest stau is predicted to be the MSSM-LSP.
Even though such a charged particle should not be the real LSP, we could consider the scenario with other neutral particle as the LSP.
One of the viable LSP candidates is gravitino, and in such a case the MSSM-LSP, $\tilde{\tau}_{1}$, possibly has a long lifetime.\footnote
{A scenario with the long-lived charged particle would conflict the cosmology.
For example, such a particle may spoil the success of the standard big-bang nucleosynthesis.
However, the problem can be evaded as long as there exists a neutral weekly interacting particle, e.g., gravitino~\cite{Kawasaki:2008qe}.
}
Indeed, the typical decay length of $\tilde{\tau}_{1}$ becomes longer than the detector size, $\sim 10$~m, if the gravitino is heavier than $O(1)$~keV.
In the following, we consider the case that we can treat $\tilde{\tau}_{1}$ as a stable particle in collider study.

We can observe the stau track when its lifetime is enough long.
This is the most striking signature of the SUSY event in the long-lived stau scenario, and it is useful for the discovery of the SUSY at the LHC~\cite{stau:cmsdiscovery}.
The main background of such an exotic charged track is muon, however, the distinction of them will be possible by using velocity and momentum information on a charged particle~\cite{Nisati:1997gb, polesello_atlmuon, Ambrosanio:2000ik, Ellis:2006vu}.
These information can be used for measuring the stau mass:
In Ref.~\cite{Ambrosanio:2000ik, Ellis:2006vu}, it is shown that the stau mass can be measured with a good accuracy $\sim O(100)$~MeV.
Stau track information is also useful for the determination of other superparticle properties and the verification of some interesting scenarios~\cite{Ambrosanio:2000ik, Ellis:2006vu, Hinchliffe:1998ys,  Hamaguchi:2004df, Feng:2004yi, Buchmuller:2004rq, Ibe:2007km, Rajaraman:2007ae, Kitano:2008en, Ishiwata:2008tp, Kaneko:2008re, Kitano:2008sa, Asai:2009ka,Feng:2009yq,  Biswas:2009zp, Ito:2009xy, Biswas:2009rba, Kitano:2010tt, Biswas:2010cd, Ito:2010xj, Endo:2010ya} because we can measure the stau's charge, velocity and momentum on event-by-event basis.

An important fact in these studies is that SUSY events in the long-lived stau scenario are thought to be background-free if a relevant velocity-cut is applied to $\tilde{\tau}_{1}$.
However, even in such a case, properties of some superparticle are hardly studied.
One of examples is the left-handed squark mass measurement.
A difficulty comes from the combinatorial background of the multiple jets.
In addition, $\tilde{q}_{L}$ has two decay modes into $q \tilde{W}^{0}$ and $q^{\prime} \tilde{W}^{\pm}$ if they are kinematically possible, therefore the number of the event of each process is small.
These problems make the measurement difficult.
Although the latter fact is problematic for the mass measurement, once we have measured $\tilde{q}_{L}$ mass in both the decay modes and confirmed two masses are identical, this is an important evidence that the squark is really the left-handed one.
In this sense, we can determine the squark chirality through the mass measurements.

In this letter, we propose methods for measuring the squark mass.
We study two sequential decay chains of the left-handed squark, $\tilde{q}_{L} \rightarrow q \tilde{W}^{0} \rightarrow q \tau \tilde{\tau}_{1}$ and $\tilde{q}_{L} \rightarrow q^{\prime} \tilde{W}^{\pm} \rightarrow q^{\prime} \nu_{\tau} \tilde{\tau}_{1}$.
In the measurements, it is important to reduce combinatorial backgrounds.
For this purpose we use a hemisphere analysis.
$\tilde{q}_{L}$ decay into $\tilde{\tau}_{1}$ is associated with a neutrino, $\nu_{\tau}$, which is produced from $\tau$ decay or $\tilde{W}^{\pm}$ decay.
The momentum reconstruction of this $\nu_{\tau}$ is another key-point and will be discussed.
We also study the right-handed squark mass measurement in the decay chain, $\tilde{q}_{R} \rightarrow q \tilde{B} \rightarrow q \tau \tilde{\tau}_{1}$.

Now we start our discussion on the squark mass measurement.
It is expected that a large number of squarks will be produced at the LHC.
Particularly, the first generation squarks, i.e., $\tilde{u}$ and $\tilde{d}$, would have large production cross sections since their superpartners are valence quarks which have large parton luminosities.
(We employ $\tilde{q}$ as either $\tilde{u}$ or $\tilde{d}$.)
A produced squark will decay to the LSP under the $R$-parity conservation.
The decay chain depends on the MSSM mass spectrum:
We study the case that the MSSM-LSP is the lightest stau and the mass spectrum in the MSSM-sector is
\begin{equation}
  m_{\tilde{q}_{R}} > m_{\tilde{B}} > m_{\tilde{\tau}_{1}},
  \label{eq:qrspetrum}
\end{equation}
\begin{equation}
  m_{\tilde{q}_{L}} > m_{\tilde{W}^{\pm}} \simeq m_{\tilde{W}^{0}} > m_{\tilde{\tau}_{1}}.
  \label{eq:qlspetrum}
\end{equation}
Such a mass spectrum is indeed realized in the minimal gauge-mediation model~\cite{Dine:1995ag, Dine:1994vc}.
In the gauge-mediation model the LSP is generally gravitino, and, as mentioned above, the stau lifetime is possibly so long.
We assume that the decay length of $\tilde{\tau}_{1}$ is much longer than the size of detectors.
For the mass spectrum in Eq.~(\ref{eq:qrspetrum}), $\tilde{q}_{R}$ decays into $\tilde{\tau}_{1}$ through the SUSY decay chain
\begin{equation}
  \tilde{q}_{R} \rightarrow q \tilde{B} \rightarrow q \tau \tilde{\tau}_{1}.
  \label{eq:qrtobino}
\end{equation}
This decay chain has a dominant branching fraction for $\tilde{u}_{R}$ and $\tilde{d}_{R}$ in general.
Although another decay mode, e.g., $\tilde{q}_{R} \rightarrow q \tilde{W}^{0}$, may be kinematically possible, such a decay only occurs through a neutralino mixing, which is predicted to be small in a variety of SUSY breaking models, and hence would have only a small branching fraction.
$\tilde{q}_{L}$ decay process is similar to the $\tilde{q}_{R}$ decay, but, for the mass spectrum in Eq.~({\ref{eq:qlspetrum}}), $\tilde{q}_{L}$ has two decay modes  because it is charged under the $SU(2)_{L}$:
\begin{equation}
  \tilde{q}_{L} \rightarrow q \tilde{W}^{0} \rightarrow q \tau \tilde{\tau}_{1},
  \label{eq:qltowino0}
\end{equation}
\begin{equation}
  \tilde{q}_{L} \rightarrow q^{\prime} \tilde{W}^{\pm} \rightarrow q^{\prime} \nu_{\tau} \tilde{\tau}_{1}.
  \label{eq:qltowinopm}
\end{equation}
The branching fraction of $\tilde{q}_{L} \rightarrow q^{\prime} \tilde{W}^{\pm}$ is approximately twice as large as that of $\tilde{q}_{L} \rightarrow q \tilde{W}^{0}$.

Measurements of masses of $\tilde{B}$, $\tilde{W}^{0}$ and $\tilde{W}^{\pm}$, which are needed in the following arguments, have been discussed in earlier works.
Several methods are proposed for the neutralino mass measurements.
For instance, when a SUSY event contains two $\tilde{B}$ and both of them decay into the $\tau \tilde{\tau}_{1}$, followed by $\tau$ decay, $\nu_{\tau}$ momenta can be reconstructed from the observed missing energy with the approximation that $\tau$ and $\nu_{\tau}$ momenta are parallel, and then $m_{\tilde{B}}$ is determined by peak analysis~\cite{Ellis:2006vu}.
Another way is to determine $m_{\tilde{B}}$ as the endpoint of the invariant mass distribution of $\tau$-jet ($j_{\tau}$) and $\tilde{\tau}_{1}$ in the decay process $\tilde{B} \rightarrow j_{\tau} \nu_{\tau} \tilde{\tau}_{1}$~\cite{Ito:2009xy}.
In the paper, it is shown that not only $m_{\tilde{B}}$ but also $m_{\tilde{W}^{0}}$ can be measured by endpoint analysis.
$\tilde{W}^{\pm}$ mass measurement has been discussed in Ref.~\cite{Biswas:2009rba}, and $m_{\tilde{W}^{\pm}}$ is determined as the endpoint of the transverse mass distribution of ($\nu_{\tau}$, $\tilde{\tau}_{1}$) system in the decay $\tilde{W}^{\pm} \rightarrow \nu_{\tau} \tilde{\tau}_{1}$.
Based on these studies, we assume in the following that $m_{\tilde{B}}$, $m_{\tilde{W}^{0}}$ and $m_{\tilde{W}^{\pm}}$ have been determined.

Squark masses can be measured by reconstructing their sequential decay processes, Eq.~(\ref{eq:qrtobino}) - Eq.~(\ref{eq:qltowinopm}).
Let us first review a method for measuring the squark mass~\cite{Ito:2009xy}.
To begin with, consider the decay of squark to the lighter stau through the on-shell neutralino, $\chi^{0}$, followed by $\tau$ decay,
\begin{equation}
  \tilde{q} \rightarrow q \chi^{0} \rightarrow q \tau \tilde{\tau}_{1}.
  \label{eq:sqtoneut}
\end{equation}
We consider only the hadronic decay mode of the $\tau$-lepton, thus, the signal consists of one jet ($j$), one $\tau$-jet and the stau.
An undetectable $\nu_{\tau}$ also accompanies to this signal.
Nevertheless, we can reconstruct the full kinematics in the process since the reconstruction of the 4-momentum of $\tau$ is possible as follows:
When $\tau$ produced in the $\chi^{0}$ decay is highly boosted, we can approximate that 4-momentum of $\tau$ and that of $\tau$-jet are collinear, that is, they can be written in the form
\begin{equation}
  p_{\tau} = r p_{j_{\tau}},
  \label{eq:colinear}
\end{equation}
where $r$ is a rescaling factor which should be in the range $r > 1$.
The value of $r$ is determined, requiring that a pair of $\tau^{\pm}$ and $\tilde{\tau}^{\mp}_{1}$ comes from the $\chi^{0}$ decay, as
\begin{equation}
  r =  \frac{ m_{\chi^{0}}^{2} - m_{\tilde{\tau}_{1}}^{2} }{ 2 p_{j_{\tau}} \cdot p_{\tilde{\tau}_{1}} }.
  \label{eq:r}
\end{equation}
We know now the kinematics in the squark decay of Eq.~(\ref{eq:sqtoneut}), completely. 
Then, by studying the distribution of the invariant mass,
\begin{equation}
  M_{ \tilde{q} } \equiv \sqrt{ ( p_{j} + r p_{j_{\tau}} + p_{\tilde{\tau}_{1}} )^2 },
  \label{eq:msq}
\end{equation}
we can determine the squark mass. Information of the mass is imprinted on the peak in the distribution. 

The above analysis can be applied to $\tilde{q}_{R}$ decay process in Eq.~(\ref{eq:qrtobino}).
By using $m_{\tilde{B}}$, the rescaling factor $r$ is determined on event-by-event basis, assuming that $\tau^{\pm}$ and $\tilde{\tau}_{1}^{\mp}$ are produced in $\tilde{B}$ decay, as
\begin{equation}
  r = r_{\tilde{B}} \equiv \frac{ m_{\tilde{B}}^{2} - m_{\tilde{\tau}_{1}}^{2} }{ 2 p_{j_{\tau}} \cdot p_{\tilde{\tau}_{1}} }.
  \label{eq:rbino}
\end{equation}
In early work~\cite{Ito:2009xy}, it has been shown that $m_{\tilde{q}_{R}}$ can been determined by this method.

One may think that the analysis is applicable to the $\tilde{q}_{L}$ mass measurement through the process in Eq.~(\ref{eq:qltowino0}), replacing $r_{\tilde{B}}$ with $r_{\tilde{W}^{0}}$, where
\begin{equation}
  r_{\tilde{W}^{0}} \equiv \frac{ m_{\tilde{W}^{0}}^{2} - m_{\tilde{\tau}_{1}}^{2} }{ 2 p_{j_{\tau}} \cdot p_{\tilde{\tau}_{1}} }.
  \label{eq:rwino}
\end{equation}
Though the event topology is quite similar between two processes, Eq.~(\ref{eq:qrtobino}) and Eq.~(\ref{eq:qltowino0}), there are some difficulties for the study of $\tilde{q}_{L}$.
SUSY event at the LHC would be characterized by multiple jets which originate from, e.g., squark decay or gluino decay.
These jets are really burdensome in the squark mass measurement because they become a source of combinatorial backgrounds.
There is no a priori way to know which jet is produced from the decay of the squark that we are interested in to measure its mass. 
To make matters worse, the decay $\tilde{q}_{L} \rightarrow q \tilde{W}^{0}$ may have a small branching fraction since there exists another decay mode into $q^{\prime} \tilde{W}^{\pm}$.
Indeed this is true, e.g., in the minimal gauge-mediation model, and the branching fraction of $\tilde{q}_{L} \rightarrow q \tilde{W}^{0}$ is roughly 30\%.
So, comparing to the right-handed squark case, the expected number of the signal is small.
For finding out such a signal we need to suppress backgrounds which mainly come from a combination of jets and also $\tau$-jets.

In order to reduce the combinatorial backgrounds, there is a useful method called hemisphere analysis~\cite{CmsTdr}.
This analysis separates reconstructed objects, such as jets, leptons and  staus, into two groups.
When the analysis successfully works, each group has constituents which arise from the decay of a heavy superparticle produced in hard-collision.
At first, we need to set a seed for each of two hemispheres. 
In the long-lived stau scenario each SUSY event contains two staus, and we assign one of staus~($\tilde{\tau}_{1}$) to the seed of the first hemisphere ($H_1$) and another stau~($\tilde{\tau}_{2}$) to the seed of the second hemisphere ($H_2$).
This assignment is quite reasonable because most of stau in SUSY event is produced from the decay of a heavy superparticle.
Next, the other objects, labeled as $i \ (=1, 2, 3, ...)$, are assigned to either of new  hemispheres, $H_1^{\prime}$ or $H_2^{\prime}$, by the criterion:
\begin{equation}
  {\rm if} \ \ d(p_{H_1}, p_i) < d(p_{H_2}, p_i),
  \ \ {\rm then} \ \ i \in H_1^{\prime},
  \ \ {\rm else} \ \ i \in H_2^{\prime},
  \label{eq:hemi}
\end{equation}
where 
\begin{equation}
  p_{H_j} \equiv \sum_{k \in H_{j}} p_k,
\end{equation}
\begin{equation}
  d(p_j, p_k) \equiv \Delta R(p_j, p_k) = \sqrt{\Delta \phi(p_j, p_k)^{2} + \Delta \eta(p_j, p_k)^{2}}.
\end{equation}
$\Delta \phi(p_j, p_k)$ and $\Delta \eta(p_j, p_k)$ are the azimuthal angle difference and the pseudo rapidity difference between $j$ and $k$, respectively.
Finally, we reassign $H_1 \equiv \{ H_1^{\prime}, \tilde{\tau}_{1} \}$ and $H_2 \equiv \{ H_2^{\prime}, \tilde{\tau}_{2} \}$, and repeat the second step, i.e., the assignment of jets and leptons.
The final step continues until the assignments to $H_j$ ($j=1, 2$) converge.

It would be useful to combine a hemisphere analysis to the mass measurement of $\tilde{q}_{L}$.
The biggest benefit is that the combinatorial backgrounds are reduced by performing the analysis on inside of each hemisphere.
As mentioned above, since multiple jets and also $\tau$-jets are expected in a SUSY event, a reduction of the combinatorial backgrounds, keeping the number of the signal 
as many as possible, will be effective.

Next we discuss the left-handed squark mass measurement by using another decay mode of $\tilde{q}_{L}$, that is, $\tilde{q}_{L} \rightarrow q^{\prime} \tilde{W}^{\pm} \rightarrow q^{\prime} \nu_{\tau} \tilde{\tau}_{1}$.
This is interesting because this analysis provides information on the left-handed squark mass, which is independent of the analysis through $\tilde{q}_{L} \rightarrow q \tilde{W}^{0} \rightarrow q \tau \tilde{\tau}_{1}$.
The signal consists of one jet, one stau and also $\nu_{\tau}$.
The momentum of $\nu_{\tau}$ from $\tilde{W}^{\pm}$ decay, denoted as $p_{\nu_{\tau}}^{(\tilde{W}^{\pm})}$, is not directly measurable but needed for the mass measurement.
For the momentum reconstruction, we consider the event
\begin{equation}
\tilde{q} \tilde{q}_{L} \rightarrow (q \chi^{0})(q^{\prime} \tilde{W}^{\pm}) \rightarrow (q \tau \tilde{\tau}_{1})(q^{\prime} \nu_{\tau} \tilde{\tau}_{1}),
\end{equation}
where $\chi^{0}$ ($\tilde{q}$) stands for either $\tilde{B}$ ($\tilde{q}_{R}$) or $\tilde{W}^{0}$ ($\tilde{q}_{L}$).
We first apply the hemisphere analysis and separates reconstructed objects into two hemispheres.
The first hemisphere is required to contain one jet, one $\tau$-jet and one $\tilde{\tau}_{1}$.
The squark reconstruction through $\tilde{q} \rightarrow q \chi^{0} \rightarrow q \tau \tilde{\tau}_{1}$ followed by $\tau \rightarrow \nu_{\tau} j_{\tau}$ has been discussed above, and the kinematics is fully known.
From Eq.~(\ref{eq:colinear}), the momentum of $\nu_{\tau}$ from $\tau$ decay is written as
\begin{equation}
  p_{\nu_{\tau}}^{(\tau)} = (r-1)p_{j_{\tau}}.
\end{equation}
Now we move on the reconstruction of the decay chain
$\tilde{q}_{L} \rightarrow q^{\prime} \tilde{W}^{\pm} \rightarrow q^{\prime} \nu_{\tau} \tilde{\tau}_{1}$.
The second hemisphere is required to contain one jet and one $\tilde{\tau}_{1}$ but no $\tau$-jets.
With the assumption that the observed missing transverse energy is equal to $(p_{\nu_{\tau}}^{(\tau)} + p_{\nu_{\tau}}^{(\tilde{W}^{\pm})})_{T}$, we obtain $x$- and $y$-components of $p_{\nu_{\tau}}^{(\tilde{W}^{\pm})}$.
Furthermore $z$-component of $p_{\nu_{\tau}}^{(\tilde{W}^{\pm})}$ is determined from the condition of
\begin{equation}
  m_{\tilde{W}^{\pm}}^{2} = ( p_{\nu_{\tau}}^{(\tilde{W}^{\pm})} + p_{\tilde{\tau}_{1}} )^{2},
  \label{eq:rec_nuwino}
\end{equation}
with a two-folding ambiguity.
In this way, the reconstruction of $p_{\nu_{\tau}}^{(\tilde{W}^{\pm})}$ is possible, but, it would be useful for a background reduction to impose the condition 
\begin{equation}
  M_{T}( p_{\nu_{\tau}}^{(\tilde{W}^{\pm})} , p_{\tilde{\tau}_{1}} ) < m_{\tilde{W}^{\pm}},
\end{equation}
where $M_{T}$ is defined by $M_{T}(p, p^{\prime}) \equiv \sqrt{ m^2 + m^{\prime 2} + 2(E_{T} E_{T}^{\prime} - {\bf p}_{T} \cdot {\bf p}_{T}^{\prime}) }$.
Then, we would gain information on the $\tilde{q}_{L}$ mass from the distribution of
\begin{equation}
  M_{ \tilde{q}_{L} } \equiv \sqrt{ ( p_{j} + p_{\nu_{\tau}}^{(\tilde{W}^{\pm})} + p_{\tilde{\tau}_{1}} )^2 }.
\end{equation}

Now, we introduce a sample MSSM-spectrum and perform a Monte Carlo analysis to demonstrate how the method described above will work.
The underlying model is assumed to be the minimal gauge-mediated model with parameters:
\begin{eqnarray}
  \Lambda = 60\ {\rm TeV}, \ \ 
  M_{\rm mess} = 900\ {\rm TeV}, \ \ 
  N_{\bf 5} = 3,\ \ 
  \tan \beta = 35,\ \ 
  {\rm sign} (\mu) = +,
\end{eqnarray}
where $\Lambda$ is the the ratio of the $F$- and $A$-components of the SUSY breaking field,
 $M_{\rm mess}$ is the mass scale of messenger fields,
 $N_{\bf 5}$ is the number of messenger multiplets in units of ${\bf 5} + {\bf \bar{5}}$ representation,
 $\tan \beta$ is the ratio of the vacuum expectation values of two Higgs bosons,
 and sign$(\mu)$ is the sign of the SUSY invariant Higgs mass parameter.
Low-energy mass spectrum is calculated by ISAJET~7.64~\cite{Paige:2003mg}, in which we use the top-quark mass of $171.3$~GeV.
Superparticle and the lightest Higgs boson masses are shown in Table~\ref{table:susymass}.
This parameter point has been studied in early work~\cite{Ito:2009xy}, and it has been shown that $\tilde{B}$, $\tilde{W}^{0}$ and $\tilde{q}_{R}$ masses are well determined.
Hence we assume that these masses have been measured.
We further assume that the mass of the lighter chargino, $\tilde{W}^{\pm}$, has been measured.
Branching fractions relevant to the following study are as follows:
 BR($\tilde{q}_{L} \rightarrow q \tilde{W}^{0}$) $\simeq$ 23\%, 
 BR($\tilde{u}_{L} \rightarrow d \tilde{W}^{+}$) $\simeq$ 49\%, 
 BR($\tilde{d}_{L} \rightarrow u \tilde{W}^{-}$) $\simeq$ 43\%,\footnote
{The soft Wino mass parameter and $\mu$-parameter are somewhat close at this parameter point.
This causes the left-handed squark decay to the heavier chargino with a considerable branching fraction.}
while $\tilde{q}_{R}$ decays into $q \tilde{B}$ with almost 100\% branching fraction.
For the lighter neutralinos, 
 BR($\tilde{B} \rightarrow \tau^{\pm} \tilde{\tau}_{1}^{\mp}$) $\simeq$ 
 BR($\tilde{W}^{0} \rightarrow \tau^{\pm} \tilde{\tau}_{1}^{\mp}$) $\simeq$ 31\%,
 and for the lighter chargino,
 BR($\tilde{W}^{+} \rightarrow \nu_{\tau} \tilde{\tau}_{1}^{+}$) $\simeq$ 61\%.
 
\begin{table}[tb]
  \centering
  \begin{tabular}{crr} 
    \hline \hline
    Particle & Mass [GeV] \\
    \hline
    $\tilde{g}    $  & 1309.39 \\
    $\tilde{u}_{L}$  & 1231.70 \\
    $\tilde{u}_{R}$  & 1183.97 \\
    $\tilde{d}_{L}$  & 1234.28 \\
    $\tilde{d}_{R}$  & 1180.19 \\
    $\tilde{t}_{1}$  & 1082.85 \\
    $\tilde{t}_{2}$  & 1195.08 \\
    $\tilde{b}_{1}$  & 1145.24 \\
    $\tilde{b}_{2}$  & 1185.83 \\
    $\tilde{\nu_{\ell}}$  & 388.05 \\
    $\tilde{\ell}_{L}$  & 396.19 \\
    $\tilde{\tau}_{2}$ & 402.57 \\
    $\tilde{\nu}_{\tau}$ & 383.80 \\
    $\tilde{e}_{R}$   & 193.39 \\
    $\tilde{\tau}_{1}$ & 148.83 \\
    $\chi^0_{1}$  & 239.52 \\
    $\chi^0_{2}$  & 425.92 \\
    $\chi^0_{3}$  & 508.41 \\
    $\chi^0_{4}$  & 548.67 \\
    $\chi^\pm_{1}$ & 425.45 \\ 
    $\chi^\pm_{2}$ & 548.43 \\
    $h$            & 115.01 \\ 
    \hline \hline
  \end{tabular}
  \caption{\small Masses of the superparticles and the lightest 
    Higgs boson $h$.}
  \label{table:susymass}
\end{table}

The SUSY event generation and the hadronization process are simulated by HERWIG~6.510~\cite{Corcella:2000bw,Moretti:2002eu}.
Events generated are passed through the fast detector simulation package, PGS~4~\cite{PGS4}, with a slight modification to treat a stable stau.
We assume that the momentum resolution of stau track is same as that of muon, and the energy deposition of stau to calorimeters is negligible.
For the stau identification, we impose two conditions:
(1)~relatively small velocity, $0.4c < v < 0.91c$;
(2)~large transverse momentum, $p_{T} > 50$~GeV.
Such staus are identified as stau, while muons are distinguished from stau under these conditions.
A stau with large velocity, $v > 0.91c$, is not identified as stau but muon.
The total SUSY cross section is $669.6$~fb in $pp$ collision with the center of mass energy of 14~TeV, and we generate all SUSY processes.
We do not consider any standard model processes as background sources;
In the long-lived stau scenario, SUSY events are characterized by the exotic slow-moving charged track of the stau.

\begin{figure}[tb]
  \centerline{\epsfxsize=0.9\textwidth\epsfbox{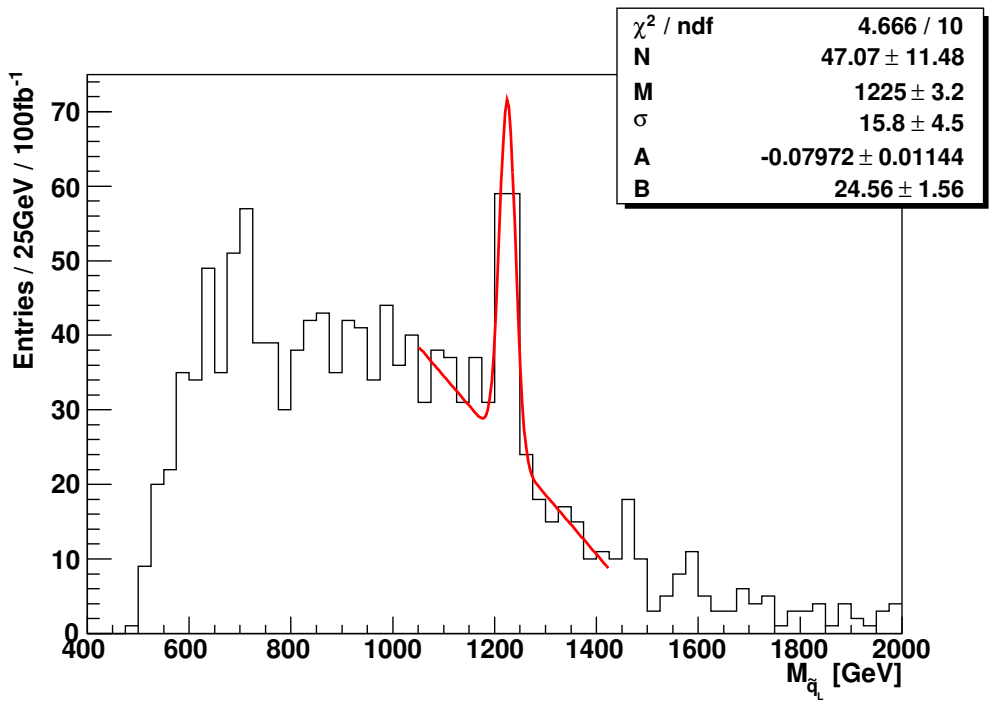}}
  \caption{\small Distribution of the invariant mass $M_{\tilde{q}_{L}} \equiv \sqrt{ ( p_{j} + r_{\tilde{W}^{0}} p_{\tau} + p_{\tilde{\tau}_{1}} )^2 }$.
  }
  \label{fig:msqwino0}
\end{figure}

We first discuss the left-handed squark mass measurement in the sequential decay process, $\tilde{q}_{L} \rightarrow q \tilde{W}^{0} \rightarrow q \tau \tilde{\tau}_{1}$.
Each of SUSY events should contain two stau tracks in the case of the long-lived  stau.
However, we require at least one stau in the signal event because $\tilde{\tau}_{1}$ with large velocity is expected to be identified as muon.
Notice that such a muon, which is a mis-tagged stau in reality, may have large $p_{T}$.
Therefore, we regard the highest $p_{T}$ muon-like track as the second stau track if there is only one slow stau in the event.
This procedure ensures that we have always two stau candidates in a SUSY event.
Then we perform the hemisphere analysis described above.
We impose two conditions so that the hemisphere method works fine:
\begin{itemize}
  \item[1A)] $\Delta R(p_{\tilde{\tau}_{1}}, p_{\tilde{\tau}_{2}}) > 2.0$, where $\tilde{\tau}_{1}$ and $\tilde{\tau}_{2}$ are the first and the second stau in the event.
  \item[1B)] $(p_{H_{j}})_{T} > 400$~GeV for $j=1,2$.
\end{itemize}
We regard objects in each of two hemispheres as daughters of a heavy superparticle produced in hard-collision.
So, for each of the hemispheres, the following conditions are imposed to find out the signal:
\begin{itemize}
  \item[1a)] Exactly one stau.
  \item[1b)] At least one $\tau$-tagged jet with $p_{T} > 30$~GeV and the charge opposite to the stau.
  \item[1c)] At least one jet with $p_{T} > 200$~GeV and $m_{\rm jet} < 80$~GeV.
  \item[1d)] No leptons with $p_{T} > 20$~GeV.
\end{itemize}
Now, we reconstruct the left-handed squark decay chain.
The momentum of $\tau$ is reconstructed from that of $\tau$-jet with the rescaling factor $r_{\tilde{W}^{0}} \ (>1)$, using $\tilde{W}^{0}$ mass information.
In this analysis, a pair of $\tau$ and $\tilde{\tau}_{1}$ from $\tilde{B}$ decay is one of backgrounds.
From Eq.~(\ref{eq:rbino}) and Eq.~(\ref{eq:rwino}), we see that the relation of $r_{\tilde{W}^{0}} > r_{\tilde{B}}$ always holds for any pair of $(j_{\tau}, \tilde{\tau}_{1})$ in the case of $m_{\tilde{W}^{0}} > m_{\tilde{B}}$.
Therefore, $\tau$-jet and $\tilde{\tau}_{1}$ from $\tilde{B}$ decay with $r_{\tilde{B}} \ (> 1)$ will be reconstructed as if they are produced in $\tilde{W}^{0}$ decay because the signal condition of $r_{\tilde{W}^{0}} > 1$ is always satisfied due to the relation of $r_{\tilde{W}^{0}} > r_{\tilde{B}}$.
In order to reduce this background we impose not only $r_{\tilde{W}^{0}} > 1$ but $r_{\tilde{B}} < 1$.
Then we study the distribution of the invariant mass
\begin{equation}
  M_{\tilde{q}_{L}} = \sqrt{(p_{j} + r_{\tilde{W}^{0}} p_{j_{\tau}} + p_{\tilde{\tau}_{1}})^{2}},
  \label{eq:mleft}
\end{equation}
where $p_{j}$ ($p_{j_{\tau}}$) is the momentum of the highest $p_{T}$ jet ($\tau$-jet).
In Fig.~\ref{fig:msqwino0}, we show the distribution of $M_{\tilde{q}_{L}}$ for the integrated luminosity ${\mathcal L} = 100$~fb$^{-1}$.
One can see the peak structure in the distribution, whose position corresponds to the squark mass.
In order to determine the squark mass, we take the fitting function, 
\begin{equation}
  f(M_{\tilde{q}_{L}}) \equiv N \exp[- (M_{\tilde{q}_{L}} - M_{\tilde{q}_{L}}^{(\rm peak)})^2 / 2 \sigma^2] + A (M_{\tilde{q}_{L}} - M_{\tilde{q}_{L}}^{(\rm peak)} )+ B,
  \label{eq:fitting}
\end{equation}
where $N, M_{\tilde{q}_{L}}^{(\rm peak)}, \sigma, A$ and $B$ are parameters.
Then, we obtain the value of $M_{\tilde{q}_{L}}^{(\rm peak)} = 1225 \pm 3.2 $~GeV while the underlying value is $m_{\tilde{q}_{L}} = 1232$~GeV.
The best-fit value is smaller than the input value.
This may be because, for example, some of jet energies are leaked from the cone used in the jet reconstruction.
Hence, such systematic errors could be corrected when the jet energy is appropriately calibrated.

\begin{figure}[tb]
  \centerline{\epsfxsize=0.9\textwidth\epsfbox{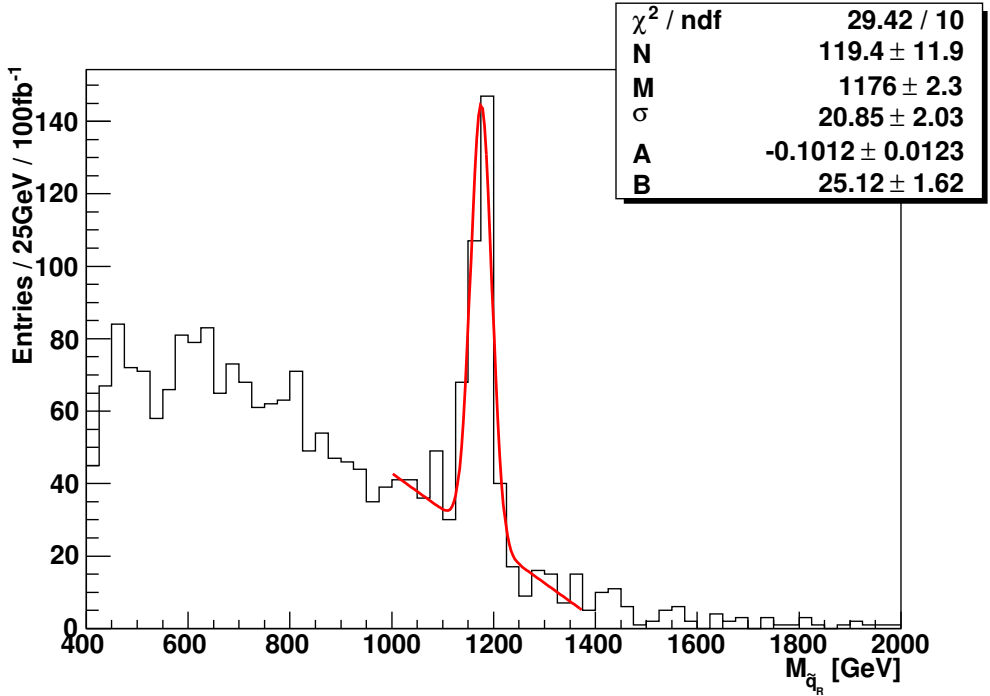}}
  \caption{\small Distribution of the invariant mass $M_{\tilde{q}_{R}} \equiv \sqrt{ ( p_{j} + r_{\tilde{B}} p_{\tau} + p_{\tilde{\tau}_{1}} )^2 }$.
  }
  \label{fig:msqbino}
\end{figure}

We also study the right-handed squark mass measurement, although it has already been discussed in early work, for a comparison.
The conditions imposed in this analysis are same in the above.
In Fig.~\ref{fig:msqbino}, we show the invariant mass distribution of $M_{\tilde{q}_{R}}$ with $r_{\tilde{B}} > 1$, where
\begin{equation}
  M_{\tilde{q}_{R}} \equiv \sqrt{ ( p_{j} + r_{\tilde{B}} p_{\tau} + p_{\tilde{\tau}_{1}} )^2 }.
  \label{eq:mright}
\end{equation}
We perform the fitting of the distribution by the fitting function introduced in Eq.~({\ref{eq:fitting}}), and obtain $M_{\tilde{q}_{R}}^{(\rm peak)} = 1176 \pm 2.3 $~GeV.
(The underlying value is $m_{\tilde{q}_{R}} = 1184$~GeV.)
Again, the small discrepancy between $m_{\tilde{q}_{R}}$ and $M_{\tilde{q}_{R}}^{(\rm peak)}$ are seen, but it could be corrected if we take some sources of systematic errors into consideration.
Comparing two figures, we see a clear difference between $M_{\tilde{q}_{R}}^{(\rm peak)}$ and $M_{\tilde{q}_{L}}^{(\rm peak)}$.
This is an important evidence that there really exist two types of squark, that is, one is the right-handed squark and the other is the left-handed squark with different masses.
Also, notice that we see from the figures the number of the $\tilde{q}_{L}$ signal is smaller than that of $\tilde{q}_{R}$ signal.
This indicates that the squark reconstructed with $r_{\tilde{W}^{0}}$ is really the $\tilde{q}_{L}$ because the smallness of the number of signal may be represented a small branching fraction of $\tilde{q}_{L} \rightarrow q \tilde{W}^{0}$.
Inversely the squark reconstructed by using $r_{\tilde{B}}$ is implied to be $\tilde{q}_{R}$.
Of course, in order to confirm the discussions here, we should  seriously consider the signal efficiencies in both analyses. 

\begin{figure}[tb]
  \centerline{\epsfxsize=0.9\textwidth\epsfbox{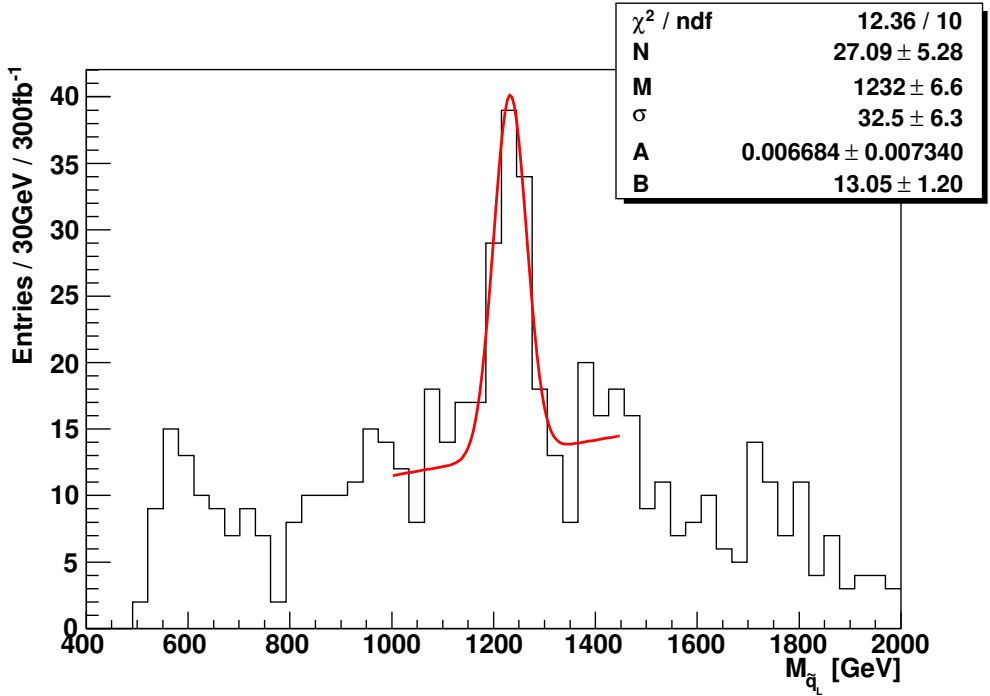}}
  \caption{\small Distribution of the invariant mass $M_{\tilde{q}_{L}} \equiv \sqrt{ ( p_{j} + p_{\nu_{\tau}}^{(\tilde{W}^{\pm})} + p_{\tilde{\tau}_{1}} )^2 }$.
  }
  \label{fig:msqchar}
\end{figure}

There exists another decay mode of $\tilde{q}_{L}$ into $q^{\prime} \tilde{W}^{\pm}$, and this decay mode can be used for the mass measurement.
We now discuss the left-handed squark mass measurement in the sequential decay process, $\tilde{q}_{L} \rightarrow q^{\prime} \tilde{W}^{\pm} \rightarrow q^{\prime} \nu_{\tau} \tilde{\tau}_{1}$.
The signal consists of one jet, one stau and also $\nu_{\tau}$ from $\tilde{W}^{\pm}$ decay.
In order to obtain the momentum of this $\nu_{\tau}$, as discussed above, we consider the event $\tilde{q} \tilde{q}_{L} \rightarrow (q \chi^{0})(q^{\prime} \tilde{W}^{\pm}) \rightarrow (q \tau \tilde{\tau}_{1})(q^{\prime} \nu_{\tau} \tilde{\tau}_{1})$.
We first use the hemisphere analysis.
One of hemispheres is required to contain:
\begin{itemize}
  \item[2a)] Exactly one stau.
  \item[2b)] At least one $\tau$-tagged jet with $p_{T} > 30$~GeV and the charge opposite to the stau.
  \item[2c)] Al least one jet with $p_{T} > 200$~GeV and $m_{\rm jet} < 80$~GeV.
\end{itemize}
If these conditions are satisfied, we calculate $r_{\tilde{B}}$ and $r_{\tilde{W}^{0}}$ from Eq.~(\ref{eq:rbino}) and Eq.~(\ref{eq:rwino}) for each pair of ($j,j_{\tau},\tilde{\tau}_{1}$).
For finding out a squark candidate, we further impose:
\begin{itemize}
  \item[2d)] If $r_{\tilde{B}} > 1$, $|M_{\tilde{q}_{R}} - M_{\tilde{q}_{R}}^{(\rm peak)} | < 40$~GeV.
  \item[2e)] If $r_{\tilde{B}} < 1$ and $r_{\tilde{W}^{0}} > 1$, $|M_{\tilde{q}_{L}} - M_{\tilde{q}_{L}}^{(\rm peak)} | < 30$~GeV.
\end{itemize}
Here $M_{\tilde{q}_{L}}$ and $M_{\tilde{q}_{R}}$ are defined in Eq.(\ref{eq:mleft}) and Eq.(\ref{eq:mright}), and we use $M_{\tilde{q}_{R}}^{(\rm peak)} = 1176$~GeV and $M_{\tilde{q}_{L}}^{(\rm peak)} = 1225$~GeV, taking into account their systematic errors.
We calculate the momentum of $\nu_{\tau}$ from $\tau$ decay by using $r_{\tilde{B}}$ if the former condition, 2d), is satisfied, or by using $r_{\tilde{W}^{0}}$  if the later condition, 2e), is satisfied.
Combining this reconstructed $\nu_{\tau}$ momentum with the observed missing transverse energy, we obtain $x$- and $y$-components of the momentum of $\nu_{\tau}$ from $\tilde{W}^{\pm}$ decay.
Now we are at the stage to reconstruct the signal process,
$\tilde{q}_{L} \rightarrow q^{\prime} \tilde{W}^{\pm} \rightarrow q^{\prime} \nu_{\tau} \tilde{\tau}_{1}$, by analyzing the other hemisphere.
We impose the conditions:
\begin{itemize}
  \item[2a)$^{\prime}$] Exactly one stau.
  \item[2b)$^{\prime}$] At least one jet with $p_{T} > 200$~GeV and $m_{\rm jet} < 80$~GeV.
  \item[2c)$^{\prime}$] No $\tau$-tagged jets with $p_{T} > 30$~GeV.
  \item[2d)$^{\prime}$] No leptons with $p_{T} > 20$~GeV.
  \item[2e)$^{\prime}$] $( p_{\nu_{\tau}}^{(\tilde{W}^{\pm})} )_{T} > 50$~GeV.
  \item[2f)$^{\prime}$] $M_{T}( p_{\nu_{\tau}}^{(\tilde{W}^{\pm})} , p_{\tilde{\tau}_{1}} ) < m_{\tilde{W}^{\pm}}$.
\end{itemize}
We reconstruct $z$-component of $p_{\nu_{\tau}}^{(\tilde{W}^{\pm})}$ by Eq.~(\ref{eq:rec_nuwino}).
Then, the squark mass is determined by the peak analysis.
As mentioned above, the two-folding ambiguity exist in this analysis, and we choose the solution which has a larger invariant mass.\footnote{
If both solutions are included in the analysis, one will see a fake peak in the distribution, which   comes from the wrong solution in the reconstruction of $p_{\nu_{\tau}}^{(\tilde{W}^{\pm})}$.
In the present parameter point, a position of the fake peak is around $M_{\tilde{q}_{L}} \simeq 1000$~GeV.
Such a fake peak will have been confirmed not to be a signal in reality, when we perform the Monte Carlo analysis using the $M_{\tilde{q}_{L}}^{(\rm peak)} \simeq 1230$~GeV and other measured superparticle masses as input parameters.
}
In Fig.~\ref{fig:msqchar}, we show the invariant mass distribution of $M_{ \tilde{q}_{L} } \equiv \sqrt{ ( p_{j} + p_{\nu_{\tau}}^{(\tilde{W}^{\pm})} + p_{\tilde{\tau}_{1}} )^2 }$, where we use the highest $p_{T}$ jet in the hemisphere.
The integrated luminosity is taken to be 300~fb$^{-1}$ in this study.
Using the fitting function given in Eq.~(\ref{eq:fitting}), we obtain $M_{\tilde{q}_{L}}^{(\rm peak)} = 1232 \pm 6.6$~GeV.
(The underlying value is $m_{\tilde{q}_{L}} = 1232$~GeV.)
Interestingly, the values of $M_{\tilde{q}_{L}}^{(\rm peak)}$ determined by two independent measurements agree with each other within the range of the statistical errors.
On the other hand, we see a significant difference between $M_{ \tilde{q}_{L} } \equiv \sqrt{ ( p_{j} + p_{\nu_{\tau}}^{(\tilde{W}^{\pm})} + p_{\tilde{\tau}_{1}} )^2 }$ and $M_{\tilde{q}_{R}} \equiv \sqrt{ ( p_{j} + r_{\tilde{B}} p_{\tau} + p_{\tilde{\tau}_{1}} )^2 }$.
This is another important indication that the squark reconstructed in the present analysis is $\tilde{q}_{L}$.
However, these results are obtained based on the simple detector simulation, and also, they are sensitive to the choice of the fitting function, the fitting range and so on.
Thus, a detailed study is needed to confirm the discussions. 

Mass measurements presented in this paper provide several information on the MSSM.
Their masses themselves reflect SUSY breaking mechanism, such as the messenger scale or the SUSY breaking scale.
After the mass measurements one could test the several SUSY breaking models by extrapolating the measured masses to high-energy scale using the renormalization group equations.
Also, superparticle properties could be determined by these analysis.
For instance, $\tilde{q}_{L}$ has two decay modes, $\tilde{q}_{L} \rightarrow q^{\prime}\tilde{W}^{\pm}$ and $\tilde{q}_{L} \rightarrow q\tilde{W}^{0}$, and
both of them can be used for the mass measurement.
If we will have confirmed that the measured masses in two analyses agrees with each other, this is an evidence that the squark is a really left-handed one.
Once squark chiralities have been determined, we can further investigate the other superparticle properties by studying a variable which reflects superparticle natures;
One of such variables is the invariant mass of $(q, \tau)$ system in the squark decay $\tilde{q}_{R} \rightarrow q \tilde{B} \rightarrow q \tau^{\pm} \tilde{\tau}_{1}^{\mp}$~\cite{Ito:2010xj}, and this variable is useful for the determination of the chiralities and spins of superparticles.

In summary, we have studied left-handed squark mass measurements in the long-lived stau scenario at the LHC experiment.
$\tilde{q}_{L}$ has, generally, two decay modes and hence the number of the signal becomes small comparing to that of $\tilde{q}_{R}$.
This is one of difficulties in $\tilde{q}_{L}$ mass measurement.
Another difficulty is a combinatorial background associated to the multiple jets and also $\tau$-jets, which is expected in the LHC experiment.
In order to reduce backgrounds and find out signals, we introduce a hemisphere analysis in which we take into account the fact that there are two stau tracks in each SUSY event if the decay length of $\tilde{\tau}_{1}$ is enough long.
We also propose some techniques to reconstruct the momentum of $\nu_{\tau}$ which is associated to $\tilde{q}_{L}$ decay.
Combining these materials, we have shown that $m_{\tilde{q}_{L}}$ is determined by peak analysis in both of two decay modes of the left-handed squark, $\tilde{q}_{L} \rightarrow q \tilde{W}^{0} \rightarrow q \tau \tilde{\tau}_{1}$ and $\tilde{q}_{L} \rightarrow q^{\prime} \tilde{W}^{\pm} \rightarrow q^{\prime} \nu_{\tau} \tilde{\tau}_{1}$, at the LHC.

\noindent
{\it Acknowledgments:}
The author would like to thank T.~Moroi for fruitful discussions and useful comments.
He also thanks R.~Kitano for the early stage contributions.


\begin{thebibliography}{99}

\bibitem{Kawasaki:2008qe}
  See, for the latest study, 
  M.~Kawasaki, K.~Kohri, T.~Moroi and A.~Yotsuyanagi,
  %``Big-Bang Nucleosynthesis and Gravitino,''
  Phys.\ Rev.\  D {\bf 78}, 065011 (2008).
  %[arXiv:0804.3745 [hep-ph]].
  %%CITATION = PHRVA,D78,065011;%%

\bibitem{stau:cmsdiscovery}
  The CMS Collaboration, CMS-PAS-EXO-08-003;
  The CMS Collaboration, CMS-NOTE-2010-008.

\bibitem{Nisati:1997gb}
  A.~Nisati, S.~Petrarca and G.~Salvini,
  %``On the possible detection of massive stable exotic particles at the LHC,''
  Mod.\ Phys.\ Lett.\  A {\bf 12}, 2213 (1997).
  %%CITATION = MPLAE,A12,2213;%%

\bibitem{polesello_atlmuon}
G. Polesello and A. Rimoldi, ATL-MUON-99-006.

\bibitem{Ambrosanio:2000ik}
  S.~Ambrosanio, B.~Mele, S.~Petrarca, G.~Polesello and A.~Rimoldi,
  %``Measuring the SUSY breaking scale at the LHC in the slepton NLSP  scenario
  %of GMSB models,''
  JHEP {\bf 0101}, 014 (2001).
  %%CITATION = JHEPA,0101,014;%%

\bibitem{Ellis:2006vu}
  J.~R.~Ellis, A.~R.~Raklev and O.~K.~Oye,
  %``Gravitino dark matter scenarios with massive metastable charged  sparticles
  %at the LHC,''
  JHEP {\bf 0610}, 061 (2006).
  %[arXiv:hep-ph/0607261].
  %%CITATION = JHEPA,0610,061;%%

\bibitem{Hinchliffe:1998ys}
  I.~Hinchliffe and F.~E.~Paige,
  %``Measurements in gauge mediated SUSY breaking models at LHC,''
  Phys.\ Rev.\  D {\bf 60}, 095002 (1999).
  %%CITATION = PHRVA,D60,095002;%%

\bibitem{Hamaguchi:2004df}
  K.~Hamaguchi, Y.~Kuno, T.~Nakaya and M.~M.~Nojiri,
  %``A study of late decaying charged particles at future colliders,''
  Phys.\ Rev.\  D {\bf 70}, 115007 (2004).
  %%CITATION = PHRVA,D70,115007;%%

\bibitem{Feng:2004yi}
  J.~L.~Feng and B.~T.~Smith,
  %``Slepton trapping at the Large Hadron and International Linear  Colliders,''
  Phys.\ Rev.\  D {\bf 71}, 015004 (2005)
  [Erratum-ibid.\  D {\bf 71}, 019904 (2005)].
  %%CITATION = PHRVA,D71,015004;%%

\bibitem{Buchmuller:2004rq}
  W.~Buchmuller, K.~Hamaguchi, M.~Ratz and T.~Yanagida,
  %``Supergravity at colliders,''
  Phys.\ Lett.\  B {\bf 588}, 90 (2004).
  %%CITATION = PHLTA,B588,90;%%

\bibitem{Ibe:2007km}
  M.~Ibe and R.~Kitano,
  %``Sweet Spot Supersymmetry,''
  JHEP {\bf 0708}, 016 (2007).
  % [arXiv:0705.3686 [hep-ph]].
  %%CITATION = JHEPA,0708,016;%%

\bibitem{Rajaraman:2007ae}
  A.~Rajaraman and B.~T.~Smith,
  %``Determining Spins of Metastable Sleptons at the Large Hadron Collider,''
  Phys.\ Rev.\  D {\bf 76}, 115004 (2007).
  %%CITATION = PHRVA,D76,115004;%%

%\cite{Kitano:2008en}
\bibitem{Kitano:2008en}
  R.~Kitano,
  %``A Clean Slepton Mixing Signal at the LHC,''
  JHEP {\bf 0803}, 023 (2008).
  % [arXiv:0801.3486 [hep-ph]].
  %%CITATION = JHEPA,0803,023;%%


\bibitem{Ishiwata:2008tp}
  K.~Ishiwata, T.~Ito and T.~Moroi,
  %``Long-Lived Unstable Superparticles at the LHC,''
  Phys.\ Lett.\  B {\bf 669}, 28 (2008).
  %%CITATION = PHLTA,B669,28;%%

%\cite{Kaneko:2008re}
\bibitem{Kaneko:2008re}
  S.~Kaneko, J.~Sato, T.~Shimomura, O.~Vives and M.~Yamanaka,
  %``Measuring Lepton Flavour Violation at LHC with Long-Lived Slepton in the
  %Coannihilation Region,''
  Phys.\ Rev.\  D {\bf 78}, 116013 (2008).
 % [arXiv:0811.0703 [hep-ph]].
  %%CITATION = PHRVA,D78,116013;%%

\bibitem{Kitano:2008sa}
  R.~Kitano,
  %``Study of chargino-neutralino production at hadron colliders in a long-lived
  %slepton scenario,''
  JHEP {\bf 0811}, 045 (2008).
  %%CITATION = JHEPA,0811,045;%%

\bibitem{Asai:2009ka}
  S.~Asai, K.~Hamaguchi and S.~Shirai,
  %``Stop and Decay of Long-lived Charged Massive Particles at the LHC
  %detectors,''
  Phys.\ Rev.\ Lett.\  {\bf 103}, 141803 (2009).
  %%CITATION = PRLTA,103,141803;%%

%\cite{Feng:2009yq}
\bibitem{Feng:2009yq}
  J.~L.~Feng, S.~T.~French, C.~G.~Lester, Y.~Nir and Y.~Shadmi,
  %``The Shifted Peak: Resolving Nearly Degenerate Particles at the LHC,''
  Phys.\ Rev.\  D {\bf 80}, 114004 (2009);
  %[arXiv:0906.4215 [hep-ph]].
  %%CITATION = PHRVA,D80,114004;%%
%\cite{Feng:2009bd}
%\bibitem{Feng:2009bd}
  J.~L.~Feng {\it et al.},
  %``Measuring Slepton Masses and Mixings at the LHC,''
  JHEP {\bf 1001}, 047 (2010).
  %[arXiv:0910.1618 [hep-ph]].
  %%CITATION = JHEPA,1001,047;%%

%\cite{Biswas:2009zp}
\bibitem{Biswas:2009zp}
  S.~Biswas and B.~Mukhopadhyaya,
  %``Neutralino reconstruction in supersymmetry with long-lived staus,''
  Phys.\ Rev.\  D {\bf 79}, 115009 (2009).
%  [arXiv:0902.4349 [hep-ph]].
  %%CITATION = PHRVA,D79,115009;%%

\bibitem{Ito:2009xy}
  T.~Ito, R.~Kitano and T.~Moroi,
  %``Measurement of the Superparticle Mass Spectrum in the Long-Lived Stau
  %Scenario at the LHC,''
  JHEP {\bf 1004}, 017 (2010).
  %%CITATION = JHEPA,1004,017;%%

%\cite{Biswas:2009rba}
\bibitem{Biswas:2009rba}
  S.~Biswas and B.~Mukhopadhyaya,
  %``Chargino reconstruction in supersymmetry with long-lived staus,''
  Phys.\ Rev.\  D {\bf 81}, 015003 (2010).
%  [arXiv:0910.3446 [hep-ph]].
  %%CITATION = PHRVA,D81,015003;%%

%\cite{Kitano:2010tt}
\bibitem{Kitano:2010tt}
  R.~Kitano and M.~Nakamura,
  %``Tau polarization measurements at the LHC in supersymmetric models with a
  %long-lived stau,''
  Phys.\ Rev.\  D {\bf 82}, 035007 (2010).
  %[arXiv:1006.2904 [hep-ph]].
  %%CITATION = PHRVA,D82,035007;%%

%\cite{Biswas:2010cd}
\bibitem{Biswas:2010cd}
  S.~Biswas,
  %``Reconstruction of the left-chiral tau-sneutrino in supersymmetry with a
  %right-sneutrino as the lightest supersymmetric particle,''
  Phys.\ Rev.\  D {\bf 82}, 075020 (2010).
  %[arXiv:1002.4395 [hep-ph]].
  %%CITATION = PHRVA,D82,075020;%%

%\cite{Ito:2010xj}
\bibitem{Ito:2010xj}
  T.~Ito and T.~Moroi,
  %``Spin and Chirality Determination of Superparticles with Long-Lived Stau at
  %the LHC,''
  arXiv:1007.3060 [hep-ph].
  %%CITATION = ARXIV:1007.3060;%%

%\cite{Endo:2010ya}
\bibitem{Endo:2010ya}
  M.~Endo, K.~Hamaguchi and K.~Nakaji,
  %``Probing High Reheating Temperature Scenarios at the LHC with Long-Lived
  %Staus,''
  JHEP {\bf 1011}, 004 (2010).
  %[arXiv:1008.2307 [hep-ph]].
  %%CITATION = JHEPA,1011,004;%%

\bibitem{Dine:1995ag}
  M.~Dine, A.~E.~Nelson, Y.~Nir and Y.~Shirman,
  %``New tools for low-energy dynamical supersymmetry breaking,''
  Phys.\ Rev.\  D {\bf 53}, 2658 (1996).
  %%CITATION = PHRVA,D53,2658;%%

\bibitem{Dine:1994vc}
  M.~Dine, A.~E.~Nelson and Y.~Shirman,
  %``Low-Energy Dynamical Supersymmetry Breaking Simplified,''
  Phys.\ Rev.\  D {\bf 51}, 1362 (1995).
  %%CITATION = PHRVA,D51,1362;%%

\bibitem{CmsTdr}
  See, for example,
  CMS Collaboration,
  ``CMS Physics TDR Vol.\ II,'' CERN/LHCC/2006-021 (2006).

\bibitem{Paige:2003mg}
  F.~E.~Paige, S.~D.~Protopopescu, H.~Baer and X.~Tata,
  %``ISAJET 7.69: A Monte Carlo event generator for p p, anti-p p, and e+ e-
  %reactions,''
  arXiv:hep-ph/0312045.
  %%CITATION = HEP-PH/0312045;%%

\bibitem{Corcella:2000bw}
  G.~Corcella {\it et al.},
  %``HERWIG 6.5: an event generator for Hadron Emission Reactions With
  %Interfering Gluons (including supersymmetric processes),''
  JHEP {\bf 0101}, 010 (2001);
%  [arXiv:hep-ph/0011363].
  %%CITATION = JHEPA,0101,010;%%
%
%\bibitem{Corcella:2002jc}
%  G.~Corcella {\it et al.},
  %``HERWIG 6.5 release note,''
  arXiv:hep-ph/0210213.
  %%CITATION = HEP-PH/0210213;%%

\bibitem{Moretti:2002eu}
  S.~Moretti, K.~Odagiri, P.~Richardson, M.~H.~Seymour and B.~R.~Webber,
  %``Implementation of supersymmetric processes in the HERWIG event
  %generator,''
  JHEP {\bf 0204}, 028 (2002).
  %%CITATION = JHEPA,0204,028;%%

\bibitem{PGS4} 
  For information on Pretty Good Simulation of high energy
  collisions (PGS4), see
  {\verb$http://www.physics.ucdavis.edu/%7Econway/research/research.html$}.

\end{thebibliography}
\end{document}